\documentstyle[prd,aps,floats,epsfig,times]{revtex}
\draft
\def\be{\begin{equation}}
\def\ee{\end{equation}}
\begin{document}
\twocolumn[\hsize\textwidth\columnwidth\hsize\csname@twocolumnfalse\endcsname
\title{Late time acceleration in Brans Dicke Cosmology}
\author{S. Sen$^{\star}$ and A. A. Sen$^{\dagger}$}
\address{Harish Chandra Research Institute, Chhatnag Road,
Jhusi. Allahabad 211
019 India}
\date{{\today}}
\maketitle
\begin{abstract}
In this work we have investigated the possibility of having a late time 
accelerated phase of the universe, suggested by recent supernova 
observation, in the context of Brans Dicke (BD) 
theory with potential having a time dependent mass squared term 
which has recently become negative and a matter field. 
We find that while a perfect fluid (pressureless and 
with pressure) cannot support this acceleration, a fluid with 
dissipative pressure can drive this late time acceleration 
for a simple power law expansion of the universe. We have also 
calculated some cosmological parameters in our model to match with 
observations.
\end{abstract}
\pacs{PACS Number(s): 04.20Jb, 98.80Hw\hfill{\sc
MRI-P-P001002}}
\vskip 2pc]

\section{Introduction}
A lot of activity has been triggered by two recent observations
\cite{Rperl,Rgar} 
on the explosion of type Ia Supernovae. These datas favour the existence 
of a new kind of matter with positive energy density dominant at present 
universe and is also responsible for the present acceleration of the 
universe accounted by its negative pressure. This along with the 
observed location of the first acoustic peak of CMB temperature 
fluctuation corroborated by the latest BOOMERANG and MAXIMA data
\cite{Rber,Rhan}, 
favours a spatially flat universe whose energy density is dominated 
by a cosmological constant like term.     
Obviously the first natural choice to represent such special matter was 
the cosmological constant $\Lambda$\cite{Rbah,Rsah}. 
For a flat matter dominated universe with $\Lambda$ having 
$\Omega_\Lambda\sim 0.72$
in Einstein gravity best fits the data sets. But the candidature of  
$\Lambda$ as the constituent of the major energy density is troubled 
by the fact that it has an energy scale which is $\sim 10^{-123}$ lower 
than normal energy scale predicted by the most particle physics models. 
So to find some alternative candidate for this acceleration a dynamical 
$\Lambda$\cite{Rcal} in the form of a scalar field with some self interacting 
potential\cite{Rpeebetc} is considered whose slowly varying 
energy density mimics 
an effective cosmological constant. The idea of this  candidate,
called {\it quintessence}\cite{Rcal}, is borrowed from the inflationary phase 
of the early universe, with the difference that it evolves at a much 
lower energy scale. The energy density of this field, though dominant 
at present epoch, must remain subdominant at very early stage and has to 
evolve in such a way that it becomes comparable with the matter density
$\Omega_m$ now. This type of specific evolution, better known as 
``{\it cosmic coincidence}''\cite{Rstei} problem, needs several constraints 
and fine tuning of parameters for the potential used to model 
quintessence with minimally coupled scalar field. A new form 
of quintessence field called ``{\it tracker field}''\cite{Rzla} has been 
proposed to solve the cosmic coincidence problem. It has an equation
 of motion with an attractor like solution in a sense that for a 
wide range of initial conditions the equation of motion converges 
to the same solution. \\

There are a number of quintessence models which have been put 
forward and most of which involve minimally coupled scalar field 
with different 
potentials dominating over the kinetic energy of the field.     
A purely exponential potential is one of the widely
studied cases \cite{Rfer}. Inspite of the other advantages 
the energy density is not enough to make up for the missing part.
Inverse power law is the other potential (\cite{Rpeebetc}-\cite{Rzla}) 
that has
been studied extensively for quintessence models, particularly for
solving the cosmic coincidence problem. Though the problems
are resolved successfully with this potential, the predicted 
value for the equation of state for the quintessence field, $\gamma_Q$,
 is not in good agreement with the 
observed results. In search of proper potentials that would 
eliminate the problems, new types of potentials, like $V_0[\cos h
\lambda\phi-1]^p$\cite{Rsah2} and $V_0\sin h (\alpha\sqrt
k_0\Delta\phi)^\beta$\cite{Rsah,Rlop} have been considered, which have
asymptotic forms like the inverse power law or exponential ones.
 Different physical considerations have lead to the study of 
other types of the potentials also\cite{Ruzan}. 
Recently Saini {\it et al}  \cite{Rsai} have reconstructed the potential 
in context of general relativity and minimally coupled quintessence 
field from the expression of the luminosity distance $d_L(z)$
as function of redshift obtained from the observational data. 
However, none of these potentials are entirely free of
problems. Hence, there is still
a need to identify appropriate potentials to explain
current observations \cite{Rfer}.\\

 Most of the studies regarding accelerated expansion 
 have been done with a minimally coupled scalar field
representing the quintessence. It has been recently shown by Pietro 
and Demaret\cite{Rpietro} that for constant scalar field equation of 
state, which is a good approximation for a tracker field solutions, 
the field equations and the conservation equations strongly constrain 
the scalar field potential, and most of the widely used potential for 
quintessence, such as inverse power law one, exponential or the cosine 
form, are incompatible with these constraints. The minimally coupled 
self interacting models will also be ruled out if the observations 
predict that the missing component of the energy density obeys 
an equation of state $p=\gamma\rho$ with $\gamma<-1 (\rho\geq 0)$ , 
and these sort of equation of state is in reasonable agreement with 
different observations \cite{Rcald}. Also the inequality 
$dH^{2}(z)/dz\geq 3\Omega_{m0}H_{0}(1+z)^{2}$ should satisfy 
for minimally coupled scalar field and its violation will certainly 
point towards a theory of non Einstein gravity such as scalar tensor 
theories where the scalar field is non minimally coupled to gravity.\\

There have been quite a few attempts
of treating this problem with the non-minimally coupled scalar fields.
 Scaling attractor solutions are available in the literature with the
exponential\cite{Ruzan} and power law\cite{Ruzan,Rlid} potentials 
in non-minimally coupled theories. Faraoni\cite{Rfar} have studied
different potentials with a non-minimal coupling term
$\psi R{\phi^2\over{2}}$ for the present acceleration. 
There have been different approach also for 
solving the problem in general scalar tensor theory, sometimes 
called {\it extended} or {\it generalised} quintessence, 
not only because this theory is considered to be the most natural 
alternative to general 
relativity, there are other strong motivations\cite{Respo} also. 
People like Bertolo {\it et al}\cite{Rbert}, Bertolami {\it et al}
\cite{Rberto},
Ritis {\it et al}\cite{Rriti} have found tracking solutions
in scalar tensor theories with different types of power 
law potential. In another work Sen {\it et al}\cite{Rsens} have found 
the potential relevant to power law expansion in Brans Dicke cosmology.
Like Saini {\it et al}\cite{Rsai}, Boisseau {\it et al} \cite{Rboi} 
have reconstructed the 
potential from the luminosity-redshift relation available from the 
observations in context of scalar tensor theory.\\

Very recently McDonald \cite{Rmcdo} has investigated the possibility 
of modelling a dynamical cosmological constant with  a scalar field 
which has undergone a very recent phase transition. For this he has 
considered a standard $\phi^{4}$ potential for the scalar field with an 
additional time dependent mass squared term in the potential which has 
become negative very recently. For this kind of model, phase transition 
occurs very recently at redshift $z\leq 1.2$.\\

In this paper we have investigated whether non minimally coupled self 
interacting scalar field such as a Brans-Dicke (BD) type scalar field  
with this kind of potential can successfully drive the late time 
acceleration for the flat universe. 
In the context of Brans Dicke(BD) 
theory\cite{Rbd} with a self interacting 
potential and a matter field, the action
is given by
\be
{\cal{S}}=\int d^4x \sqrt{-g}[\phi
R-{\omega\over{\phi}}\phi^\alpha\phi_\alpha
-V(\phi)+{\cal{L}}_m]\label{action}
\ee
(We have chosen the unit $8\pi G_0=c~=~1$.)\\
We have chosen the $\phi^{4}$ potential with a time dependent mass 
squared term which has already become negative after a phase transition in 
recent time \cite{Rmcdo}: 
\be
V(\phi)=\lambda \phi^4-\mu^2(t)\phi^2
\label{pot}
\ee  
where 
\be
\mu^2(t)=\bar{\mu_0}^2\left(\frac{R_c}{R}\right)^n=\frac{\mu_0^2}{R^n}
\ee
${\mu_0^2}=\bar{\mu_0}^2{R_c}^n$. $\lambda$ is a constant and $n$ is an 
integer. The time dependent mass squared term with integer $n$ can 
arise naturally in plausible models and one can have a detailed discussion in \cite{Rmcdo}. \\

As a matter field we would consider first perfect fluid and 
then a fluid having negative pressure. An effective negative pressure 
and hence an acceleration can be achieved by dissipative mechanism 
modelled commonly by fluid viscosities. It has been proposed recently
that the CDM must self interact in order to explain the detailed 
structure of the galactic halos\cite{Rsperg}. This self interaction 
will naturally create a viscous pressure whose magnitude will depend 
on the mean free path of the CDM particles. An effective negative 
pressure in CDM can also be created from cosmic anti friction which is 
closely related to particle production out of gravitational field\cite{Rzim}.
Since the negative pressure 
can be modelled in different ways, we are not apriori assuming 
any specific model for this negative pressure.\\

In this work we find that
it is not possible to have a late time power law accelerated expansion 
when the CDM is a perfect fluid, but a dissipative CDM fluid in 
BD cosmology with such a potential like (\ref{pot}) can successfully 
drive a late time accelerated expansion.
In the next section we treat the field equations and find the solutions 
in both the cases. We also calculate some cosmological parameters to match 
the accelerated model with observation. The third section is the 
concluding section where 
we have discussed different features of this model. 

\section{Field Equations and Solutions}
The gravitational field equations derived from the action (\ref{action})
by varying the action with respect to the metric is,
\begin{eqnarray} 
G_{\mu\nu} = \frac{T_{\mu\nu}}{\phi}+\frac{\omega}{\phi^2} (\phi_\mu \phi_\nu
-\frac{1}{2}g_{\mu\nu}\phi_\alpha\phi^\alpha)\nonumber
\\
+\frac{1}{\phi}[{\phi_\mu}_{;\nu}
-g_{\mu\nu}\Box{\phi}]-g_{\mu\nu}\frac{V(\phi)}{2\phi}
\label{fldeqn}
\end{eqnarray}
where $T_{\mu\nu}$ represents the energy momentum tensor of the matter 
field. We have assumed the matter content of the universe to be composed 
of a fluid represented by the energy momentum tensor 
\be
T_{\mu\nu}=(\rho+P){\it{v}}_\mu{\it{v}}_\nu+P
g_{\mu\nu},\label{emtensor}
\ee
where $\rho$ and $P$ are the energy density and effective pressure of 
the fluid respectively and  ${\it v}_\mu$is the four velocity of the fluid
i.e, ${\it{v}}_\mu{\it{v}}^\mu=-1$. The effective pressure of the fluid 
includes the thermodynamic pressure $p$ and a negative pressure $\pi$, 
which could arise either because of the viscous effect or due to particle 
production, i.e, 
\be
P=p+\pi
\label{pressure}
\ee
The wave equation that follows from equation (\ref{action}), by varying 
the action with respect to the scalar field $\phi$ is
\be
\Box\phi=\frac{T}{2\omega+3}+\frac{1}{2\omega+3}\left(\phi\frac{dV(\phi)}
{d\phi}-2V(\phi)\right)\label{waveeqn}
\ee 
For our choice of potential (\ref{pot}), the field equations (\ref{fldeqn}) 
and the wave equation (\ref{waveeqn}) becomes respectively
\be
3{\dot R^2\over{R^2}}+3{\dot
R\over{R}}{\dot\phi\over{\phi}}-{\omega\over{2}}{\dot\phi^2\over{\phi^2}}-
\frac{\lambda}{2}\phi^3+\frac{\mu_0^2}{2R^n}\phi={\rho\over{\phi}},
\label{fldeqn1}
\ee
\be
2{\ddot{R}\over{R}}+{\dot R^2\over{R^2}}+{\ddot
{\phi}\over{\phi}}+2{\dot
R\over{R}}{\dot\phi\over{\phi}}+{\omega\over{2}}{\dot\phi^2\over{\phi^2}}-
\frac{\lambda}{2}\phi^3+\frac{\mu_0^2}{2R^n}\phi =-{(p+\pi)\over{\phi}}
\label{fldeqn2}
\ee
and
\be
\ddot{\phi}+3\frac{\dot R}{R}\dot{\phi}=\frac{\rho-3(p+\pi)}{2\omega+3}-\frac{1}
{2\omega+3}\left[2\lambda\phi^4+\frac{n\mu_0^2\phi^2}{R^n}\frac{\frac{\dot R}{R}}
{\frac{\dot\phi}{\phi}}\right]
\label{wveqn}
\ee
We have assumed standard Friedman-Robertson-Walker metric with the 
signature convention $(-,+,+,+)$ and $R$ is the scale factor. 
We restrict ourselves for spatially
flat metric only. We work in Jordan frame. One interesting thing about 
BD theory in Jordan frame is that the conservation equation holds for 
matter and scalar field separately. Or in a slightly different way, the 
Bianchi Identity along with the wave equation (\ref{waveeqn}) gives 
the matter conservation equation
\be
\dot\rho+3{\dot R\over{R}}(\rho+p+\pi)=0
\label{coneqn}
\ee
We assume that both the scale factor and the scalar field evolve as 
the power function of time
\be
R=R_0 \left({t\over{t_0}}\right)^\alpha~~~~{\rm{and}}~~~~
\phi=\phi_0 \left({t\over{t_0}}\right)^\beta \label{Rphi}
\ee
 where the subscript $0$ refers to the values
of the parameters at the present epoch. In order to get accelerated 
expansion for such a evolution of the universe the deceleration parameter 
has to be negative, which immediately restricts the parameter $\alpha$
to be greater than 1. For such an expansion the solution 
for the matter density is
\be
\rho = \rho_c t^{\beta-2}  \label{rho}
\ee
where 
\be
\rho_c = {3\alpha\phi_0\over{t_o^\beta}}
\left[{2\alpha+\beta(1+\alpha)-\beta^2(1+\omega)\over{2-\beta}}\right]
\label{rhoc}
\ee
First we consider normal perfect fluid with no negative pressure i.e, 
$\pi=0$ in equation (\ref{pressure}). Then, for the power law 
evolution, the thermodynamic pressure of the fluid becomes
\be
p = p_c t^{\beta-2} \label{p}
\ee
where 
\be
p_{c} = {(2-\beta-3\alpha)\phi_0\over{t_o^\beta}}
\left[{2\alpha+\beta(1+\alpha)-\beta^2(1+\omega)\over{2-\beta}}\right]
\label{pc}
\ee
Power law solution is consistent with the field equations 
(\ref{fldeqn1}), (\ref{fldeqn2}) and (\ref{wveqn}) only if 
\be
\beta=-\frac{2}{3}~~~{\rm and}~~~\alpha n-\beta=2~~~{\rm i.e,}~~
\alpha n=\frac{4}{3}\label{value}
\ee
So the acceleration demands $n<\frac{4}{3}$, and as $n$ is an positive integer, it restricts $n=1$ in our model. Again from equation (\ref{rho})
the weak energy condition $(\rho>0)$ demands 
\be
\omega<3\alpha-\frac{5}{2}             .
\label{17a}
\ee
 From equation (\ref{p}) 
and (\ref{rho}) it is clear the perfect fluid follows an equation of state 
of the form $p=\gamma_m\rho$. The index $\gamma_m$ is given by
\be
\gamma_m={2-\beta\over{3\alpha}}-1, \label{gammam}
\ee
 where $\gamma_m$ lies within the interval $0<\gamma_m<1$. 
This restricts $\alpha$ within the
range $\frac{4}{9}<\alpha<\frac{8}{9}$. Infact for present matter 
dominated universe $(\gamma_m=0)$, $\alpha=\frac{8}{9}$. But this does
not satisfy the criteria for acceleration $(\alpha>1)$ and hence 
for a simple power law type expansion the universe decelerates with 
a perfect fluid CDM ($0\leq\gamma_{m}<1$) with a potential (\ref{pot}) 
in BD theory.

Now we consider a CDM which has a dissipative effect and we are 
particularly interested in a present day universe i.e, $p=0$. Under such
condition equation (\ref{coneqn}) takes the form
\be
\dot\rho+3{\dot R\over{R}}(\rho+\pi)=0
\label{coneqn2}
\ee
As is mentioned earlier this type of dissipative effect in FRW
cosmology can be modelled in two ways. Generally the dissipative effect 
is accounted by conventional bulk viscous effect. In FRW universe the 
bulk viscosity can be modelled within the framework of non-equilibrium
thermodynamics proposed by Israel and Stewart\cite{Ris}. According to 
this theory the bulk viscous pressure $\pi$ follows the transport equation
\be
\pi+\tau\dot{\pi}=-3\eta H-\frac{\tau\pi}{2}[3H+\frac{\dot{\tau}}{\tau}
-\frac{\dot{T}}{T}-\frac{\dot{\eta}}{\eta}]
\label{transporteqn}
\ee
where the positive definite quantity $\eta$ stands for the coefficient
 of bulk viscosity, $T$ is the temperature of the fluid and $\tau$ is 
the relaxation time associated with the dissipative effect i.e, the 
time taken by the system to reach equilibrium state if the dissipative 
effect is suddenly switched off. Considering the divergence term in 
the square bracket to be small i.e, $(\frac{R^3\tau}{\eta T})$ to be 
constant, the equation can be approximated to a simpler form
\be  
\pi+\tau\dot{\pi}=-3\eta H
\label{truntraneqn}
\ee
In literature this is commonly described as a truncated version of the 
full nonequilibrium thermodynamics. The viscous effects are assumed to be
not so large as observation seems to rule out huge entropy production on 
large scales\cite{Rstec}. Usually $\tau$ is expressed as $\frac{\eta}{\rho}$
so as to ensure that the viscous signal does not exceed the speed of 
light\cite{Rbel} and also $(\tau H)^{-1}=\nu$, where $\nu>1$ for a 
consistent hydrodynamical description of the fluid\cite{Rcoley}. With these 
two assumption equation (\ref{truntraneqn}) becomes
\be
\nu H+\frac{\dot{\pi}}{\pi}=-\frac{3\rho H}{\pi}.
\label{vis}
\ee
In a very recent work Chimento et 
al\cite{Rchim} have shown that a mixture of minimally coupled self 
interacting scalar field and a perfect fluid is unable to drive 
accelerated expansion and solve the cosmic coincidence problem at 
the same time, while the mixture of a dissipative CDM with bulk viscosity 
along with minimally coupled self interacting scalar field can 
successfully drive the accelerated expansion and solve the cosmic 
coincidence problem simultaneously.\\

An effective negative pressure can also be created from cosmic 
antifriction which is closely related to particle production out
of gravitational field. In a recent paper Zimdahl {\it et al}\cite{Rzim}
have shown that one can have a negative $\pi$ if there exists a 
particle number nonconserving interaction inside matter. This may 
happen due to particle production out of gravitational field. In 
this case, the matter is of course not a dissipative fluid, but a 
perfect fluid with varying particle number. Though substantial 
particle production is an event that occurs in early universe, 
Zimdahl {\it et al} have shown that extremely small particle production 
rate can also cause sufficiently negative $\pi$ to violate strong 
energy condition.\\  

We do not apriori assume any specific model for this dissipative effect, 
rather only assume the existence of a negative $\pi$. 
For a similar kind of evolution of the scale factor and the scalar field
given by equation (\ref{Rphi}), the energy density for the fluid, 
with negative pressure is also given by equations (\ref{rho}) and
(\ref{rhoc}).  From equations (\ref{coneqn2}) and (\ref{rho}), 
one can easily find that
\be
\pi={(2-\beta-3\alpha)\phi_0\over{t_o^\beta}}
\left[{2\alpha+\beta(1+\alpha)-\beta^2(1+\omega)\over{2-\beta}}\right]
t^{\beta-2}
\label{pi}
\ee
From equation (\ref{pi}) one can easily check that to have a negative 
$\pi$, one should have  $3\alpha>2-\beta$ 
which essentially means $\alpha>\frac{8}{9}$. This suits the condition 
for acceleration as an $\alpha$ is needed to be greater than 1 for that.
One can also check that for this set of solutions given by 
(\ref{Rphi}), (\ref{rho}) and (\ref{pi}) 
and from equations (\ref{vis}), 
the condition $\nu>1$ holds provided 
$2-\beta>0$, which is very much true in our case. This is important 
for the hydrodynamical description if the CDM is assumed to be a 
conventional viscous fluid.\\

To have a clear picture of the expansion of the universe and the 
missing energy, we further 
study the energy density and pressure of the geometric scalar field. 
The expressions for the energy density and the pressure of the scalar 
field can be derived from the field equations (\ref{fldeqn1}) and 
(\ref{fldeqn2}) to be
\be
\rho_\phi=
\left[{\omega\over{2}}~{\dot\phi^2\over{\phi}}+{V\over{2}}-3{\dot
R\over{R}}\dot\phi\right]
\label{rhop}
\ee
and
\be
p_\phi=\left[{\omega\over{2}}~{\dot\phi^2\over{\phi}}-{V\over{2}}+\ddot\phi
+2{\dot R\over{R}}\dot\phi\right]
\label{pphi}
\ee
In case of power law expansion (\ref{Rphi}) and potential like (\ref{pot})
the energy density of the BD field becomes
\be 
\rho_\phi=\frac{\alpha\phi_0}{2t_0^\beta} \left\{3\alpha+\omega+\frac{5}{2}
\right\}
t^{\beta-2}
\label{rhop1}
\ee
and pressure of the BD field is
\be
p_\phi=\left[-\frac{\alpha}{2}\left(3\alpha+\omega+\frac{5}{2}\right)
+\frac{2}{3}\left(\alpha+\frac{2}{3}\omega+\frac{5}{3}\right)\right]
\frac{\phi_0}{t_0^\beta}t^{\beta-2}
\label{pphi2}
\ee
The positivity condition for the scalar energy density demands
\be
\omega>-(3\alpha+\frac{5}{2})
\label{27a}
\ee
 which eventually  
restricts $\omega$ beyond some lower value. So 
essentially the two positivity energy conditions,(\ref{17a}) and (\ref{27a}), 
limit the range of 
$\omega$ within $-(3\alpha+\frac{5}{2})<\omega<3\alpha-\frac{5}{2}$.  
Clearly a barotropic relation $(p_\phi=\gamma_\phi\rho_\phi)$ is 
followed by the scalar field, where the adiabatic index $\gamma_\phi$
is given by
\be  
\gamma_\phi=~-1+\frac{\frac{2}{3}\left(\alpha+\frac{2}{3}\omega+
\frac{5}{3}\right)}
{\frac{\alpha}{2}\left(3\alpha+\omega+\frac{5}{2}\right)}
\label{gamma}
\ee
The range of $\gamma_\phi$ that agrees with the observational datas 
and describes the current acceleration for the universe well, is 
$-0.6>\gamma_\phi>-1$. One can adjust the value of $\alpha$ and $\omega$
so as to get the required value of $\gamma_\phi$. We
now recast equation (\ref{fldeqn1}) in the form  
\be
\Omega_m+\Omega_\phi=1
\label{densityparameter}
\ee
where the density parameters for matter $\Omega_m$ and scalar field 
$\Omega_\phi$ are defined to be (see ref \cite{Rdiaz})
\be
\Omega_m=\frac{\rho}{3H^2\phi}~~~{\rm and}~~~\Omega_\phi=
\frac{\rho_\phi}{3H^2\phi}
\label{omegamphi}
\ee
The expression for density parameters at present epoch are
\be
\Omega_{m0}=\frac{\rho_0}{3H_0^2\phi_0}=\frac{1}{2}-\frac{1}{6\alpha}
(\omega+\frac{5}{2})
\label{omegam0}
\ee
and
\be
\Omega_{\phi 0}=\frac{\rho_{\phi 0}}{3H_0^2\phi_0}=\frac{1}{2}+
\frac{1}{6\alpha}
(\omega+\frac{5}{2})
\label{omegap0}
\ee
Like $\gamma_\phi$, the value of $\Omega_{m0}$ that suits best the 
luminosity distance-redshift data for type Ia supernovae is 
$\Omega_{m0}=0.28$ and in a similar fashion like $\gamma_\phi$, 
one can adjust $\alpha$ and $\omega$ value to get the required 
value of $\Omega_{m0}$ that tally with the observation.\\

So far $\alpha$ is restricted only by the deceleration parameter
that it should be greater than 1 for the universe 
to accelerate. And the positivity energy conditions limits $\omega$ 
within the range $-(3\alpha+\frac{5}{2})<\omega<3\alpha-\frac{5}{2}$.
So if a small value is chosen for $\alpha~(\sim 1)$, $\omega$ is also 
restricted accordingly and a suitable choice of both parameters can be 
made to find the allowable range of  $\gamma_\phi$ and $\Omega_{m0}$ 
that matches observation. 
Kaplinghat et al\cite{Rkapling} and others\cite{Rsethi}
 have pointed out that for 
power law cosmologies, high redshift data and present age of the universe 
restricts $\alpha$ to a value $\approx 1$. In a very recent 
investigation we have found that the best fit value of $\alpha$ with 
the SN Ia data for power law cosmology is approximately $1.25$\cite{Rsen}. 
Hence this small value for $\alpha$ restricts $\omega$ also to be small.
 But this squarely contradicts the solar system bound on 
$\omega(>600)$. To accommodate such large value of $\omega$, $\alpha$ 
should be large.  But for the large values of $\alpha$ the 
universe accelerates faster ( almost like de-sitter expansion)
and $\gamma_{\phi}$ asymptotically approaches $-1$. It is quite 
unlikely that the universe presently accelerates in such a high 
power law fashion and such $\alpha$ values do not match with present 
observation. We would discuss more about this point in the discussion 
section.\\

We wish to find the range of the parameters $\alpha$ and $\omega$ allowable 
in our model , that suits the permissible range of $\gamma_\phi$ and 
$\Omega_{m0}$ of the quintessence proposals. 
In figure 1, we have shown the allowed region in the $(\alpha,\omega)$ 
parameter space (shaded portion in the figure) for the specified range 
of $\gamma_\phi$ ($-0.6>\gamma_\phi>-0.8$) and $\Omega_{m0}$ 
($0.5>\Omega_{m0}>0.3$), where we have assumed $\alpha$ is small. It can 
be noticed that the allowable range of 
both the parameters $\alpha$ and $\omega$ obeys the constraints, 
(\ref{17a})and (\ref{27a}), imposed 
on them by physical conditions.

\begin{figure}[hb]
\centering
\leavevmode\epsfysize=7cm \epsfbox{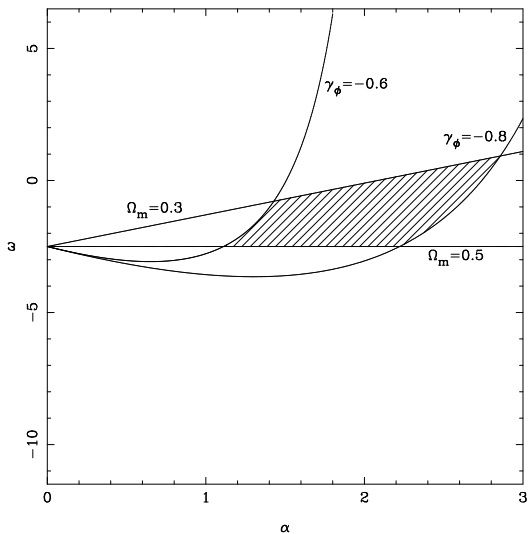}
\caption{$\omega$ {\it vs} $\alpha$ for $-0.6>\gamma_\phi>-0.8$ and 
$0.5>\Omega_{m0}>0.3$}
\label{figure1}
\end{figure}

A point to note here is that in BD theory the gravitational coupling 
$G$ varies inversely with the scalar field $\phi$. At present time 
$\phi$ approaches a constant value $\phi_0$, the inverse of which 
gives the present Newtonian constant $G_N$. In the weak field limit 
the present Newtonian coupling and the asymptotic value of $\phi$ 
is related by 
\be
G_N=\frac{2\omega+4}{2\omega+3}~~\frac{1}{\phi_0}
\label{phi0}
\ee
 The present day variation of the gravitational coupling $G$, 
is $\frac{\dot{G}}{G}|_0=\frac{2}{3\alpha}H_0$, where 
$H_0~(=\frac{\alpha}{t_0})$ is the Hubble parameter at present. 
For any value of $\alpha$ that allows acceleration
this rate is $<10^{-10}$ per year\cite{Rwill}.

Another important point to mention here is that time variation of $G$ does not 
directly affect the nuclear process of the early universe. But the 
expansion rate of the universe in  this type of theory do influence the 
primeval nucleosynthesis\cite{Rwein}.
A fixed value for the parameter $\alpha>1$ for all epochs implies that 
universe 
is always accelerating which seriously contradicts the nucleosynthesis 
scenario. One way to avoid such problem is to consider $\omega$ as a 
function of the scalar field $\phi$. In a recent work Banerjee and Pavon
\cite{Rban} have shown that with $\omega(\phi)$ one can have a decelerating 
radiation dominated era in the early time and accelerated matter dominated 
era in the late time. But in their case also, $\omega$ has to be small 
asymptotically to have a late time acceleration for the universe.

To analyse the nature of acceleration and our ansatz more critically, 
it is interesting to match different cosmological parameters with 
observations. We intend to find the age of the universe and the 
luminosity distance-redshift relation compatible with our model, 
probing the background dynamics, that could differentiate between 
different types of universe.\\

Since one of the main incentive for reconsidering the introduction 
of cosmological constant was the age of the universe, we first consider 
the age of the universe suggested in our model and the constraints 
imposed on it by observations.
Equation (\ref{fldeqn1}) can also be presented as 
\be  
H^2=H_0^2[\Omega_{m0}+\Omega_{\phi 0}](1+z)^{\frac{2}{\alpha}}
\label{hsquare}
\ee

\begin{figure}[hb]
\centering
\leavevmode\epsfysize=7cm \epsfbox{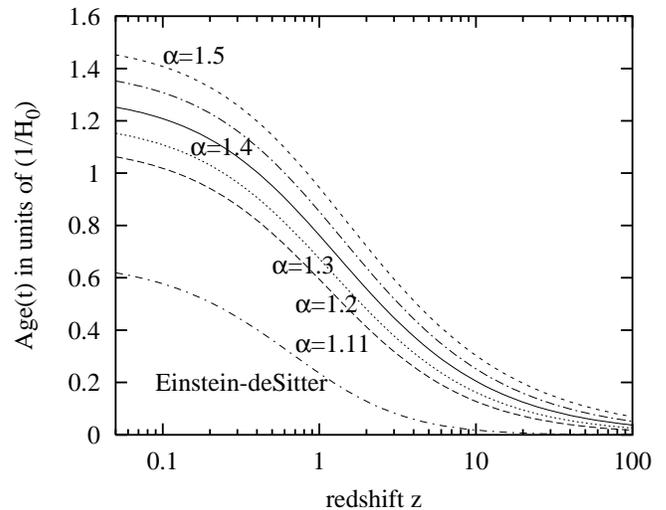}
\caption{Age $(t)$ {\it vs} redshift $z$ for different $\alpha$ values}
\label{figure2}
\end{figure}

where $z$ is the redshift defined by 
\be
1+z=\frac{R_{observed}}{R_{emitted}}
\label{redshift}
\ee

From equation (\ref{hsquare}) we find the age of the observable universe
for a given redshift $z$, is
\be
t_0-t=\frac{\alpha}{H_0(\Omega_{m0}+\Omega_{\phi 0})^{\frac{1}{2}}}\left[
1-\frac{1}{(1+z)^{\frac{1}{\alpha}}}\right]
\label{aget}
\ee
Of course for $t=0$, i.e, for infinite redshift, the age of the universe 
is $t_0={\frac{\alpha}{H_0}}$.\\

An old object observed at a certain redshift selects all models with 
at least that age at that given redshift. In that respect, several 
age constraints have recently appeared in the literature\cite{Rdulop}.
For example, the age of the radio galaxy 53W091 observed at a redshift
 $z=1.55$ puts a lower bound of 3.5 Gyrs at that redshift. The quasar 
observed at $z=3.62$ sets a lower bound of 1.3 Gyrs. In figure 2 we 
present a plot of the age of the universe as a function of redshift for 
various values of $\alpha$. Taking into account the range of $\alpha$,
prescribed by figure (\ref{figure1}), our universe has an age limit of 
$t_0\geq 15.5Gyrs$. From the figure it can be seen that all the universes
with that minimum age limit , except that of Einstein-desitter one, 
always accommodate these constraints. Recently Pont {\it et al}\cite{Rpont} 
estimated the age of the universe to be $14\pm 2$ Gyrs, 
which is in excellent agreement with our result.\\

Now we would like to trace the change of luminosity distance with respect
 to the redshift in our model so as to compare it with the present data 
available. The result that reveal the so called acceleration of the 
universe was the observation of the luminosity distance as a function
of redshift for type Ia supernovae, which is believed to be a standard 
candle. From almost 60 redshifts, 42 high redshift data obtained by 
Supernova Cosmology Project and 18 low redshift observed by Calan Tololo 
Supernova Survey, favour a universe with positive cosmological constant.
Assuming flatness in context of general relativity, the best fit for 
these data occurs for $\Omega_{m0}=0.28$  and $\Omega_{\lambda 0}=0.72$.
Optical astronomers measure luminosities in logarithmic units, called 
magnitudes, given by
\be
m(z)={\cal M}+5\log d_L+25
\ee
where ${\cal M}$ is the absolute magnitude and $d_L$ is luminosity 
distance defined by

\be
d_L=R(t_0)(1+z)r_1
\ee
for an event at $r=r_1$ at time $t=t_1$.\\
According to our ansatz the 
expression for $d_L$ is
\be
d_L(z)=\frac{(1+z)}{H_0(\Omega_{m0}+\Omega_{\phi 0})^{\frac{1}{2}}}
\int^{z}_0 F(z\prime)d{z\prime}
\ee
where $F(z)=\frac{1}{(1+z)^{\frac{1}{\alpha}}}$. \\

In figure 3 we have 
plotted this luminosity distance versus redshift for different values of 
$\alpha$. We see that for different $\alpha$ values, the $d_L$ is 
practically same for lower redshifts upto $z\sim~0.4$. At redshifts 
$z>0.4$ the curves are separated, but is not distinctly separate  
to discriminate and rule out different types of the models. Therefore, 
high accuracy measurements with uncertainties at percentage level
are needed in order to cleanly distinguish the models and 
the need to go 
to redshifts sensibly higher than 1 is evident. In this respect it is 
very much relevant to mention that Supernova Acceleration Probe (SNAP) 
is planned to make measurements with an accuracy at percentage level 
upto redshifts $z\sim1.7$\\
 
\begin{figure}[hb]
\centering
\leavevmode\epsfysize=7cm \epsfbox{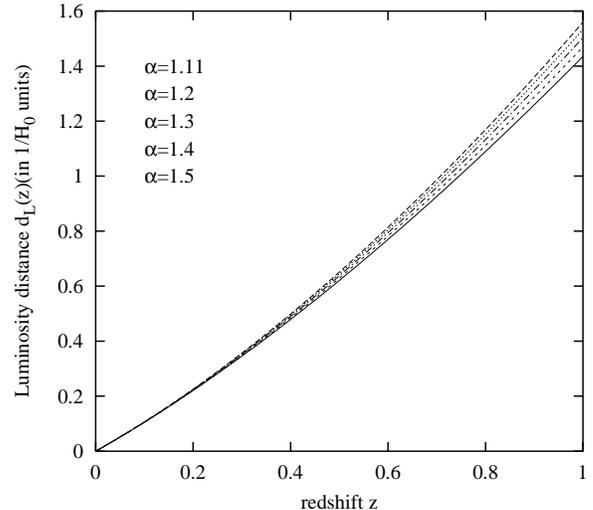}
\caption{Luminosity distance $(d_L)$ {\it vs} redshift $z$ for different $\alpha$ values}
\label{figure3}
\end{figure}

\section{Discussion}
This work investigates the possibility of getting an accelerated 
universe in context of BD theory with a $\phi^{4}$ potential having a time dependent mass squared term  and a matter field. In this work we have not used quintessence 
field to trace the missing energy. The BD scalar field , which is a 
geometric scalar field, plays the role of dynamical $\Lambda$ and 
provides that missing energy. It is found that for a simplistic 
approach of power law expansion ($\sim t^{\alpha}$)
a perfect fluid kind of matter (both pressureless and with pressure) 
cannot support a late time acceleration of the universe, if the  
scalar field have potential given by equation (\ref{pot}).   
But a matter with a dissipative effect can provide the acceleration 
that agrees with the observational data sets. The dissipative effect 
accounted by the negative pressure can be modelled in two ways according 
to the recent investigations\cite{Rzim,Rchim}. Particle production out 
of gravitational field can give rise to negative pressure while energy 
can also be dissipated by bulk viscous effect between the CDM particles. 
 In this work though no particular model is considered for the origin of the 
negative pressure, it is found that $\nu>1$. This is important for 
hydrodynamical description if the CDM is assumed to be a conventional 
dissipative fluid. We have also calculated 
different parameters like the time variation of gravitational coupling, 
age of the universe  and the luminosity-distance redshift relation.
All of these cosmological parameters agrees quite well with the recent 
observations.\\
 
The accelerated solution depends 
crucially upon two parameters: $\alpha$ and BD parameter $\omega$, both 
of which are constrained by different physical conditions. Different 
combinations of $\alpha$ and $\omega$ can produce the required values for 
$\gamma_\phi$ ($-0.6>\gamma_\phi>-1$) and $\Omega_m$  ($\sim0.3$)
 that tallies with present observation suggesting acceleration.
Small $\alpha$ values restricts $\omega$ to small negative values and 
support the late time acceleration scenario quite successfully.The 
cosmological parameters calculated with the small value of $\alpha$ 
agrees with observations quite well. Many references
\cite{Rkapling,Rdiaz,Rsethi} are available in 
the literature where it has been shown that this value should be 
very close to $1$ to be consistent 
with observation. In one of our recent work\cite{Rsen} it is 
shown that for simple power law expansion $(\sim t^\alpha)$ of the 
universe, the best fit value of $\alpha$ with SNIa data is approximately 
1.25, and due to (\ref{17a}) and (\ref{27a}), this will restrict $\omega$ 
to small value. But this clearly contradicts the solar system limit 
$\omega>600$. One should note that in our model, large value of 
$\omega$ consistent with solar system limit,  is not restricted 
either by physical conditions such as positivity of energy density 
or by the requirements of specific ranges for $\gamma_{\phi}$ or 
$\Omega_{m}$ consistent with the observations. Only the fact, that 
$\alpha$ is not large, is  
 the prediction of the datas obtained so far and this
constrains $\omega$ to small value. If future observations predicts 
large value for $\alpha$ then that can also be accommodated in our 
model with a large value for $\omega$.\\

One should also note that, for $\alpha<1.33$ which is consistent with 
the present data, the dissipative pressure is not sufficient to drive 
the acceleration alone, and the BD scalar field along with the 
dissipative pressure in the CDM drives the acceleration whereas the 
BD scalar field plays the role of the missing components of the universe.\\

It is also expected that after the phase transition, one expects $\phi$ to 
roll down from $\phi=0$ taking different values in different directions 
causing large scale inhomogeneities. But it was argued by McDonald 
\cite{Rmcdo} that after this recent phase transition the universe will 
be filled by non topological objects like ``{\it axions}'' whose radius 
$r_{\phi}$ is much smaller than $10 Mpc$ causing these axions to behave 
like a smooth dark energy components.\\

The range of $\omega$, we obtained in our calculations in order to fulfil 
different physical conditions and also to have correct range of values for 
$\gamma_{\phi}$ , $\alpha$, and $\Omega_{m}$, is consistent with the range 
obtained by other authors \cite{Rsens,Rsen,Rban}. Also, it can be seen in 
the figure 3 that for different values of $\omega$ and $\alpha$, it is 
difficult to distinguish between models  upto $z\sim 1$. Hence our model 
is not very much fine tuned as far as the parameter $\alpha$ and $\omega$ 
are concerned. \\

It is also important to note that as $\alpha$ remains constant throughout 
the age of the universe. This essentially means that 
the universe always accelerates for $\alpha>1$ , which 
seriously contradicts the primeval nucleosynthesis scenario. As we 
have mentioned earlier that one way to overcome this problem is to 
consider $\omega$ to be function of the scalar field $\phi$
\cite{Rban}. A choice of $\omega$ (polynomial function of $\phi$) 
can give a decelerating radiation era as well as accelerating matter 
dominated era. But then also $\omega$ asymptotically acquires a small 
negative value for an accelerating universe at late time. 
 In most 
of the  investigations done in scalar tensor theory\cite{Rsens,Rsen,Rban} 
such conclusions 
have been arrived. Only exception, so far in our knowledge, is that done by 
Bertolami and Martins\cite{Rberto}, where in Brans Dicke cosmology with 
a $\phi^2$ 
potential, the solution of an accelerated universe is obtained with large 
$|\omega|$. But there the positive energy condition on both matter and 
the the scalar field have not been considered. There are also other 
evidences in literature where small $|\omega|$ has been supported.
In the extended inflationary model by La and Steinhardt ~\cite{LS}, 
the required value for $\omega$ is 20. 
The structure formation in scalar tensor theory also contradicts 
the solar system bound on $\omega$\cite{Rgaz}.Thus the problem 
seems to appear in different scales (astronomical and cosmological). 
The theory has been tested by experiments so far only in the 
astronomical scales and no experiment had been done in cosmological 
scale as yet. And so the problem occurs in finding the compatibility 
between astronomical observation and cosmological requirements. 
Considering $\omega$ to be a variable to have both decelerating and 
accelerating phases at different epochs, while large $\omega$ values 
occur due to local inhomogeneities in astronomical scale to satisfy 
the solar system bound, can be a complete investigation and may 
give a satisfactory answer to this question.

\end{document}